\documentclass[fleqn,10pt,a4paper]{wlscirep}
\usepackage{graphicx,amsmath,amssymb,wasysym,verbatim,color}

\title{Competing contagion processes: Complex contagion triggered by simple contagion}
\author[1,2$*$]{Byungjoon Min}
\author[1,$\dagger$]{Maxi San Miguel}
\affil[1]{IFISC, Instituto de F\'isica Interdisciplinar y Sistemas Complejos
(CSIC-UIB), Campus Universitat Illes Balears, E-07122 Palma de Mallorca, Spain}
\affil[2]{Department of Physics, Chungbuk National University, Cheongju, Chungbuk 28644, Korea}
\affil[*]{bmin@chungbuk.ac.kr}
\affil[$\dagger$]{maxi@ifisc.uib-csic.es}
\date{\today}

\begin{abstract}
Empirical evidence reveals that contagion processes often occur with competition of
simple and complex contagion, meaning that while some agents follow simple
contagion, others follow complex contagion. Simple contagion refers to spreading
processes induced by a single exposure to a contagious entity
while complex contagion demands multiple exposures for transmission.
Inspired by this observation, we propose a model of contagion dynamics with 
a transmission probability that initiates a process of complex contagion. 
With this probability nodes subject to simple contagion get adopted and 
trigger a process of complex contagion. We obtain a phase diagram in the parameter 
space of the transmission probability and the fraction of nodes subject to complex 
contagion. Our contagion model exhibits a rich variety of phase transitions such
as continuous, discontinuous, and hybrid phase transitions, criticality,
tricriticality, and double transitions. In particular, we find a double phase
transition showing a continuous transition and a following discontinuous transition
in the density of adopted nodes with respect to the transmission probability.
We show that the double transition occurs with an intermediate phase in which nodes
following simple contagion become adopted but nodes with complex contagion
remain susceptible.
\end{abstract}

\begin{document}
\flushbottom
\maketitle
\thispagestyle{empty}
\maketitle

\section*{Introduction}

Models of social and biological contagion in general fall into two classes depending
on the response to successive exposures: simple and complex contagion
\cite{goffman,daley,schelling,granovetter,may,watts,castellano,centola2,macy,centola,karsai,vespignani}.
Simple contagion, mainly inspired by disease spreading, stands for a contagion process
with independent interaction between the susceptible and the infectious~
\cite{may,castellano,kermack,pastor,newman,epidemic,layer,vazquez,muhua}.
Typical compartment epidemic models such as the susceptible-infected-recovered model~
\cite{may,kermack,newman,epidemic} and the susceptible-infected-susceptible~
\cite{may,pastor,epidemic} model belong to the class of simple contagion processes. Models of simple
contagion are controlled by an infection probability independent of
the number of exposures.
Typically, a model of simple contagion exhibits a continuous phase transition between an
epidemic phase and a disease free phase for a critical value of the infection probability. 
The other class of contagion processes is complex contagion representing spreading
phenomena in which multiple exposures to a spreading entity are needed for changing
agents' state \cite{centola2,macy}. Models of complex contagion processes encompass a
wide range of contagious models such as threshold model \cite{granovetter,watts,dodds},
generalized epidemic model \cite{janssen,chung,choi1,choi2,gomez}, diffusion
percolation \cite{adler}, threshold learning \cite{avella,lugo},
and bootstrap percolation \cite{chalupa}. The spread of fads, ideas, and new
technologies in our society is better described by complex
contagion rather than by simple contagion due to a collective effect in social
contagion \cite{weng,lerman,monsted,kramer,hodas,voter}. The critical difference
of the complex contagion as compared to the simple contagion processes is that the probability
of adoption depends on the number of exposures. For instance, in the threshold
model the adoption of a new innovation
happens when the number of adopted neighbors is larger than a certain
threshold \cite{granovetter,watts}. Models of complex contagion often result in
a discontinuous phase transition in contrast to the continuous phase transition
of simple contagion~\cite{watts,ruan,gleeson-seed,brummitt,lee,voter}.

Classical contagion models assume that the contagious entity determines
the type of contagion either simple or complex \cite{epidemic,weng}.
Recently, the comprehensive analysis of the spread of
an equal-sign profile in a social networking service (SNS) \cite{dow,state,kara}
sheds light on the mechanisms of contagion processes between the susceptible and the adopted.
The analysis of the empirical data shows that characteristics of the agents also
affects the type of contagion \cite{state}. Specifically, it is observed that the
number of successful exposures requiring for adoption is far different for
different individuals \cite{state}. In this observation, some agents change their
profile in SNS just after the first exposure to the meme (simple contagion), but the
others need more exposures to be adopted (complex contagion).
This implies a competition between simple and complex contagion depending
on agents' adoptability, deviating from the traditional view of contagion models.
The heterogeneity of adoptability can be widespread for many spreading phenomena
because of the individual diversity of stubbornness, creed, and preexisting information.
These facts call for incorporating such heterogeneity when modelling contagion processes
integrating simple and complex contagion \cite{bottcher,aga,cellai,baxter,wang}.
Incorporating such heterogeneity, here we propose a contagion model that in addition 
considers a transmission probability in the contagion process
acting like an infection probability in epidemic models or an occupation
probability in bond percolation processes on a network. It represents a trial
of transmission from adopted neighbors, prior to the subsequent adoption processes.
Effectively the transmission probability acts as a simple contagion process triggering 
a process of complex contagion

In our model of contagion processes with
a transmission probability we unify simple and complex contagion by considering agents with heterogeneous
adoptability. We assign explicitly a different level of adoptability
for individuals to mimic the heterogeneity of adoptability observed in empirical
data \cite{state}.
The transmission probability models a chance to transmit and to identify successful 
(active) connections for adoption processes.
With these generalizations, our model includes a variety of contagion models such as
the susceptible-infected-recovered model~\cite{kermack},
threshold model \cite{granovetter,watts}, diffusion percolation \cite{adler},
and bootstrap percolation \cite{baxter2}.
Our generalized contagion model exhibits a rich variety of phenomena including
continuous, discontinuous, and hybrid phase transitions, criticality, tricriticality,
and double transitions. We show that a double transition with an intermediate phase
can happen when a system is composed of nodes with heterogeneous adoptability.

\section*{Generalized contagion model}

\begin{figure}
\includegraphics[width=\linewidth]{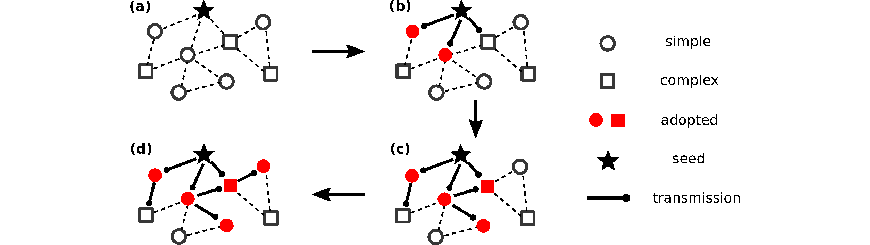}
\caption{
An example of our model of contagion processes with a transmission probability
unifying simple and complex contagion. In this example, five nodes (circles) out
of nine nodes follow simple contagion and three nodes (squares) follow complex
contagion requiring two exposures to be adopted.
Spreading starts from a seed (star symbol) and susceptible nodes (open symbols) become
adopted (filled symbols) when the number of successful exposures exceeds or equates 
its assigned adoptability either $1$ for simple contagion or $2$ for complex contagion.
}
\end{figure}

We consider a network with $N$ nodes that can be in a susceptible or adopted state. 
The adoptability $\theta$ of each node
is randomly drawn from a distribution $Q(\theta)$. To be specific, $\theta$
represents the number of successful exposures required to change from susceptible to adopted.
For example, when $\theta=1$, a node becomes adopted after a single successful
exposure thus indicating simple contagion, while when $\theta>1$, it represents
complex contagion node since multiple exposures are needed for adoption. Varying the
adoptability $\theta$, we can describe both simple and complex contagion processes.
The chance of transmission is determined by a transmission probability $\lambda$.
Each adopted node attempts to spread with the probability $\lambda$ and hence
we can identify active connections between the susceptible and the adopted.
Introducing the distribution of adoptability $Q(\theta)$ and the transmission
probability, we unify the two contagion mechanisms and suggest a generalized
contagion model. It is worthwhile to note that heterogeneous adoptability
but without a transmission probability was considered in the threshold
model \cite{granovetter,kara}, heterogeneous $k$-core
percolation \cite{cellai,baxter}, and a model of stochastic
interacting particles \cite{raul}.

In our model, dynamics is in discrete time. Initially, all nodes are 
susceptible except a fraction $\rho$ of seed nodes that are adopted.
Newly adopted nodes attempt transmission with a probability $\lambda$
to all of their susceptible neighbors in the same time step. In the next
time step, each susceptible node updates the number of successful transmissions
and becomes adopted if the number of successful exposures is
the same or larger than its threshold. 
In more detail: let us suppose that an adopted node $i$ tries
transmission to its susceptible neighbor $j$ with a probability $\lambda$.
If the transmission is successful, the link from $i$ to $j$ becomes active
and with the complementary probability $1-\lambda$, the link remains inactive.
Then, susceptible node $j$ becomes adopted when the number of successful exposures
(equivalently the number of active links towards node $j$)
exceeds or equates its adoptability $\theta_j$. This process proceeds until there are no
more newly adopted agents in a network.

The main parameters of our model are $\lambda$ and $Q(\theta)$ which reflect respectively
the extent of transmissibility of a contagious entity and the adoptability distribution of the nodes.
Depending on these two parameters, our model becomes one of a wide range of contagion models. 
The susceptible-infected-recovered model \cite{kermack,newman} is recovered when 
($\lambda,Q(\theta)$)=($\lambda,\delta_{\theta,1}$) where $\delta_{i,j}$ represents the
Kronecker delta function (the function is $1$ if $i=j$ and $0$ otherwise).
Diffusion percolation \cite{adler} corresponds to ($\lambda,Q(\theta)$)=($1,\delta_{\theta,n>1}$) where $n$
is any integer larger than unity, while Watts' threshold model \cite{watts} corresponds to
($\lambda,Q(\theta)$)=($1,\delta_{\theta,k_i T}$) where $T$ is a threshold and $k_i$ is the degree of node $i$.

Figure 1 shows an example of our generalized contagion model with
$Q(\theta)=(1-p)\delta_{\theta,1} + p \delta_{\theta,2}$. A fraction $(1-p)$ of nodes
denoted by circles follow simple contagion ($\theta=1$) and a fraction $p$ of nodes denoted by
squares follow complex contagion requiring multiple successful exposures to become adopted ($\theta>1$).
Initially, all nodes are susceptible except a seed indicated with a star symbol [Fig.~1(a)].
Next, adopted nodes attempt to spread the contagious entity with a probability $\lambda$. If a trial
is successful, a susceptible node is exposed to a contagious entity (denoted by thick line).
Note that a single success of transmission does not always result in
adoption because complex contagion requires multiple successful exposures.
When the number of successful exposures exceeds or equates the adoptability $\theta$ of a node,
a susceptible node turns to the adopted state (filled symbols) [Fig.~1(b-d)]. Eventually
we measure the final fraction of adopted nodes $R$ at the steady state.

\section*{Analytical approach}

To predict the final fraction of adopted nodes, we derive mean-field
equations assuming a locally tree-like structure in the limit $N \rightarrow \infty$.
Our approximation is exact in a tree structure and it gives very good agreement 
with numerical simulations for sparse random graphs with only infinite loops.
Our approach is based on recent theoretical developments for the threshold 
cascade model on networks \cite{gleeson-seed}.
A generating function technique developed for a model of percolation processes \cite{newman}
also shares the idea of our analytical treatment. Given a degree distribution $P(k)$ and
an adoptability distribution $Q(\theta)$, the expected final fraction of adopted nodes
$R$ from a fraction of initial seed nodes $\rho$ (chosen randomly) can be expressed
as \cite{gleeson-seed,newman},
\begin{align}
\label{eq:r}
R = &\rho+ (1-\rho) \sum_{k=0}^{\infty} P(k) \sum_{m=0}^{k} \binom{k}{m} q_{\infty}^m (1-q_{\infty})^{k-m}
\sum_{\theta=1}^{\infty} Q(\theta) \left[ 1 - \sum_{s=0}^{\theta-1} \binom{m}{s} \lambda^s (1-\lambda)^{m-s} \right].
\end{align}
Here $q_{\infty}$ is the steady state probability that a node is adopted
by following a randomly chosen link, and $\lambda$ is the transmission probability
between the susceptible and the adopted.
The term $\binom{k}{m} q_{\infty}^m (1-q_{\infty})^{k-m}$ corresponds to the probability
of having $m$ adopted neighbors out of $k$ neighbors. And,
$\left[1 - \sum_{s=0}^{\theta-1} \binom{m}{s} \lambda^s (1-\lambda)^{m-s} \right]$
represents the probability that the number of successful exposures with the transmission probability
$\lambda$ from $m$ adopted neighbors exceeds or equates the adoptability $\theta$.
Overall, Eq.~\ref{eq:r} corresponds to the probability that a randomly chosen node is either a seed
node with probability $\rho$ or is not a seed with the probability $(1-\rho)$
but it becomes eventually adopted in the dynamical process.

The probability $q_{\infty}$ can be obtained by solving a recursive equation.
First we define $q_t$ as the probability that a node is adopted
by following a randomly chosen link at level $t$.
On a locally tree-like graph, $q_t$ can be obtained by 
\begin{align}
\label{eq:q}
q_{t+1} = &\rho+ (1-\rho) \sum_{k=1}^{\infty} \frac{k P(k)}{\langle k \rangle} \sum_{m=1}^{k-1} \binom{k-1}{m} q_t^m (1-q_t)^{k-m-1}
\sum_{\theta=1}^{\infty} Q(\theta) \left[ 1 - \sum_{s=0}^{\theta-1} \binom{m}{s} \lambda^s (1-\lambda)^{m-s} \right].
\end{align}
The fixed point of the above equation corresponds to $q_{\infty}$ starting from
the initial value $q_0=\rho$.
In general, we obtain $q_{\infty}$ by solving iteratively Eq.~\ref{eq:q} and
obtain $R$ by replacing the value obtained for $q_{\infty}$ in Eq.~\ref{eq:r}.

We further develop the theory for an Erd\"os-R\'enyi (ER) graph with an average
degree $z$ as a simple example. ER graphs in the limit
$N \rightarrow \infty$ clearly satisfy the locally tree-like structure, and
hence our theoretical calculation gives a good approximation.
Using the degree distribution $P(k)=e^{-z} z^k/k!$,
the final fraction of adopted nodes $R$ becomes the same as $q_{\infty}$
since Eq.~\ref{eq:r} and \ref{eq:q} become equivalent.
Then, the self-consistency equation is simply expressed as
\begin{align}
\label{eq:er}
R & = \rho+(1-\rho) \sum_{\theta=1}^{\infty} Q(\theta) \left[ 1-e^{- z \lambda R} \sum_{i=1}^{\theta} \frac{( z \lambda  R)^{i-1}}{(i-1)!}\right] \nonumber \\
 & = \rho+(1-\rho) \sum_{\theta=1}^{\infty} Q(\theta) \left[ 1- \frac{\Gamma(\theta, z \lambda R)}{\Gamma(\theta)} \right],
\end{align}
where $\Gamma(x)$ is the gamma function and $\Gamma(x,y)$ is the incomplete gamma function.
Thus, for ER networks, we can obtain the fixed point of $R$ directly by solving the above
self-consistency equation.

\section*{Results}

\subsection*{Phase diagram}

For the sake of simplicity, we consider a model with a bimodal distribution of the adoptability
$Q(\theta)=(1-p)\delta_{\theta,1} + p \delta_{\theta,n}$ on ER networks.
In this setting, a fraction $(1-p)$ of nodes follows simple contagion with $\theta=1$
(simple nodes) and a fraction $p$ of nodes follows complex contagion requiring
$n$ successful exposures to be adopted (complex nodes). We then have three parameters,
$p$, $n$, and $\lambda$ which respectively correspond to the fraction of complex nodes,
the number of successful exposures required for complex nodes to adopt, and the 
probability of transmission. Further assuming that the initial density of seed nodes is
negligible, i.e, $\rho \rightarrow 0$, the self-consistency equation becomes,
\begin{align}
\label{eq:two}
R &= (1-p)\left(1-e^{-z \lambda R}\right)+p \left[1-\frac{\Gamma(n, z \lambda R)}{\Gamma(n)} \right].
\end{align}
The first term corresponds to the contribution of simple nodes
and the second term corresponds to that of complex nodes.

In order to identify a fixed point of Eq.~\ref{eq:two}, we define
$f(R)=-R+(1-p)\left(1-e^{-z \lambda R}\right)+p\left[1-\frac{\Gamma(n, z \lambda R)}{\Gamma(n)} \right]$.
Then, the fixed points $R^*$ are given by the zeros of $f(R^*)=0$.
We find that the trivial solution $R^*=0$ indicates adoption free phase where
adoption does not happen both for simple and complex nodes.
Adoption phase showing non-zero density of adopted nodes ($R>0$) appears 
at the point where the trivial solution $R^*=0$ becomes unstable.
Linear stability analysis implies that the adoption free phase is stable 
when $f'(0)<0$ while it becomes unstable if $f'(0)>0$.
Thus, the transition between the adoption free
phase ($R=0$) and the adoption phase ($R>0$) occurs at $f'(0)=0$ where
$f'(R)=-1+(1-p)(z \lambda) e^{-z \lambda R}+ p \frac{(z \lambda)^n R^{n-1}}{\Gamma(n)}e^{-z \lambda R}$.
From the condition $f'(0)=0$, we obtain the transition point $\lambda_1$
for any positive integer $n$,
\begin{equation}
\lambda_1^{(n,p)}=
\begin{cases}
\quad \frac{1}{z} & \text{if } n=1, \\
\quad \frac{1}{z(1-p)} & \text{if } n>1.
\end{cases}
\end{equation}
When $n=1$, all nodes are simple nodes meaning that the model returns to an ordinary
simple contagion, i.e., essentially the same as the susceptible-infected-recovered model \cite{kermack}.
Therefore, the threshold for simple contagion model $1/z$ is recovered \cite{newman}.
When $n>1$, we get an additional $(1-p)$ factor which corresponds to the fraction of simple nodes.

\begin{figure}
\includegraphics[width=\linewidth]{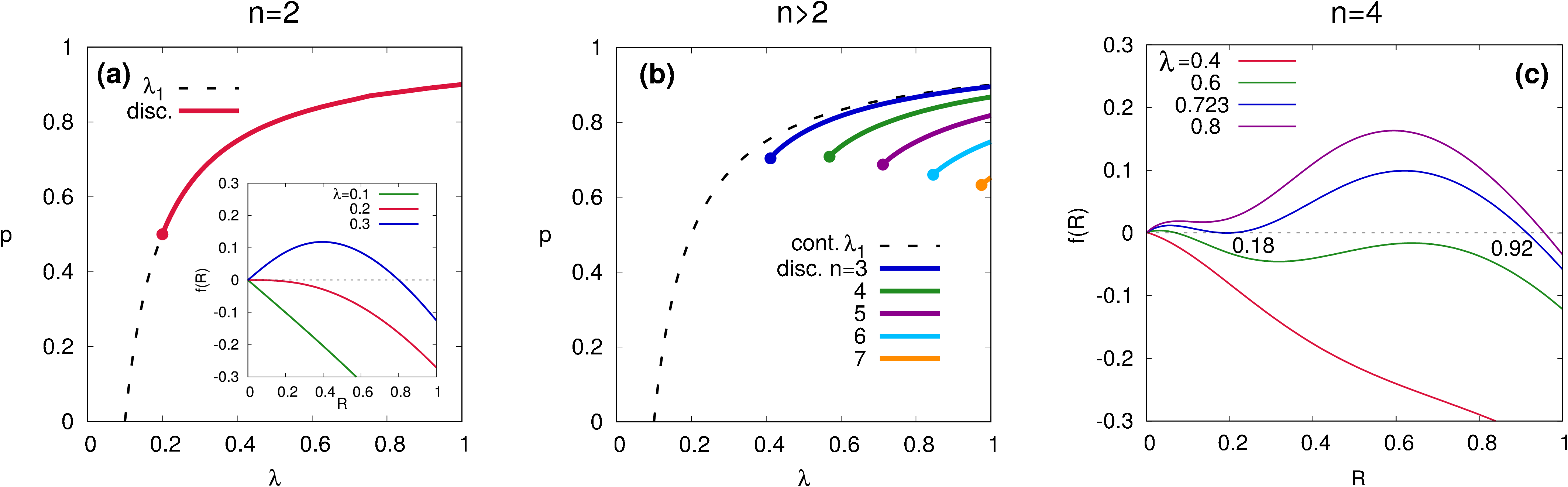}
\caption{
Phase diagram of a generalized contagion model with (a) $n=2$ and (b) $n>2$ for ER
networks with $z=10$. Continuous and discontinuous transition lines are respectively
indicated as dashed and solid lines, and (tri)critical points are indicated by filled circles.
Graphical solution of $f(R)$ at $p=p_{tc}=1/2$ and $\lambda=0.1,0.2(\lambda_{tc}),0.3$
is shown in the inset of (a).
(c) Graphical solution of $f(R)$ with $n=4$, $p=0.8$ and
$\lambda=0.4,0.6,0.723(\lambda_2),0.8$ is shown.
}
\end{figure}

The nature of the transition at $\lambda_1$ is determined by the second derivative of $f(R)$.
While the transition is continuous if $f''(0)<0$, it becomes
discontinuous if $f''(0)>0$. Applying this condition to
$f''(R)=-(1-p)(z \lambda)^2 e^{-z \lambda R} -  p (1-n+ z \lambda R) \frac{(z \lambda)^n R^{n-2}}{\Gamma(n)}  e^{-z \lambda R}$,
we find that $f''(0)<0$ for all values of $p$ if $n>2$. Therefore the transition at
$\lambda_1$ is always continuous if $n>2$. However, when $n=2$, $f''(0)<0$ for
$p<0.5$ and $f''(0)>0$ for $p>0.5$, so that the transition is continuous
if simple nodes hold a majority ($p<0.5$) and discontinuous if complex nodes
hold a majority ($p>0.5$). In this case of $n=2$, we can further identify a tricritical point
($\lambda_{tc},p_{tc}$)=($2/z,1/2$) by imposing the condition $f''(0)=f'(0)=f(0)=0$
where the continuous and discontinuous transition lines meet.
At the tricritical point, the size of the discontinuous jump for $p<0.5$ reduces to zero.

In the phase diagram with $n=2$ for ER networks with $z=10$ [Fig.~2(a)], we find continuous
(dashed) and discontinuous (solid) transition lines and a tricritical point
$(\lambda_{tc},p_{tc})=(0.2,0.5)$ at which the two lines meet [Fig.~2(a)].
For $p<p_{tc}$, the transition at $\lambda_1$ is continuous with the scaling behavior
$R\sim (\lambda-\lambda_1)^{\beta_1}$ and the exponent $\beta_1=1$, 
the same as the mean-field exponent of an ordinary bond percolation \cite{newman}.
Approaching the tricritical point, we obtain a different scaling
$R \sim (\lambda-\lambda_{tc})^{\beta_{tc}}$ with $\beta_{tc}=1/2$. 
For $p>p_{tc}$, the transition at $\lambda_1$ becomes discontinuous.
In the inset of Fig.~2(a), the graphical solution $f(R)$ with respect to $R$
at $p=p_{tc}$ with $\lambda=0.1,0.2,0.3$ is shown.
The zeroes of $f(R)$ correspond to the fixed point and
$\lambda=0.2$ corresponds to the tricritical point $\lambda_{tc}$
in our example with $z=10$.
In the adoption free phase, there exists only a trivial solution which is $R^*=0$.
When $\lambda$ is larger than the tricritical point value ($\lambda>\lambda_{tc}$),
a new stable solution appears at a non-zero value of $R^*$ and $R^*=0$ solution becomes unstable.

For $n>2$, in addition to the continuous transition at $\lambda_1$ with 
the critical exponent $\beta_1=1$ for all $n>2$,
there is another transition at $\lambda_2^{(n,p)}$
which is discontinuous, indicated by a solid line [Fig.~2(b)].
It is worthwhile to note that $\lambda_2$ is larger than $\lambda_1$ for any $n>2$.
The location of $\lambda_2$ can be analytically identified from the
condition $f'(R^*)=0$ with $R^* \ne 0$.
When $n>2$, the continuous transition line $\lambda_1$ and the discontinuous line
$\lambda_2$ are separated and do not meet. Thus, the tricriticality at which
the continuous and discontinuous
transition lines meet is a peculiar behavior only found in $n=2$.
The size of the discontinuous jump at $\lambda_2$ decreases with decreasing $p$ and goes
to zero at a critical point $(\lambda_c,p_c)$ indicated by a filled circle,
at which $f''(R^*)=f'(R^*)=0$.
Thus, the discontinuous transition line ends at the critical point
and there is no the second phase transition when $p<p_c$.
In this regime ($p<p_{c}$), $R$ increases gradually without discontinuity
when increasing $\lambda$ with $\lambda>\lambda_1$.
In addition, the discontinuous jump and critical point can disappear 
as $n$ increases for a given $z$, i.e, for $z=10$, there is no second transition
when $n>7$.

When $p>p_c(n)$ with $n>2$, the adoption phase is separated into two distinct phases
by a boundary at $\lambda_2$: simple adoption (low $R$) and complex adoption (high $R$) phases.
In addition, the transition at $\lambda_2$ has hybrid characteristics showing both
discontinuity and a scaling behavior, $R(\lambda_2)-R \sim (\lambda_2-\lambda)^{\beta_2^{(n>2)}}$
with the exponent $\beta_2^{(n>2)}=1/2$ for any $n>2$. 
In addition, when $\lambda$ approaches the critical point a cube-root scaling appears as
$R-R(\lambda_c) \sim (\lambda-\lambda_c)^{\beta_c}$ where $\beta_c=1/3$. 
This is the same scaling found in heterogeneous $k$-core percolation \cite{baxter}.
Such hybrid phase transition also known as mixed phase transition has been observed
widely in cooperative percolation in networks such as $k$-core
percolation \cite{baxter,goltsev,dlee}, bootstrap percolation \cite{baxter2},
percolation of interdependent networks \cite{buldyrev,baxter-aval,viability},
and cooperative epidemic processes \cite{chung,gomez,choi1,choi2}.
A hybrid transition is also predicted in a model of spin chains with long-range
interactions \cite{thouless}, DNA denaturation \cite{mukamel}, and
jamming \cite{schwarz,sheinman}, and recently observed experimentally
in a colloidal crystal \cite{alert}.

An example of how to identify a phase transition is shown in terms of the graphical solution 
of $f(R)$ with $z=10$ and $n=4$ in the limit $\rho \rightarrow 0$ [Fig.~2(c)]. 
The zeroes of $f(R)$ give the fixed point values of $R$ and their stability is given by 
the derivative of $f(R)$. First, $R$ remains zero for $\lambda<\lambda_1$.
When $\lambda_1<\lambda<\lambda_2$, $R$ increases gradually
as $\lambda$ increases until the second transition at $\lambda=\lambda_2$.
As $\lambda$ increases further $\lambda>\lambda_2$, a complex adoption phase ($R\approx 0.92$)
appears suddenly from the simple adoption phase ($R \approx 0.18$). Therefore, our analysis
predicts the emergence of a double transition showing a continuous and 
a subsequent discontinuous transition.

\subsection*{Continuous, discontinuous, and double phase transitions}

\begin{figure}
\includegraphics[width=\linewidth]{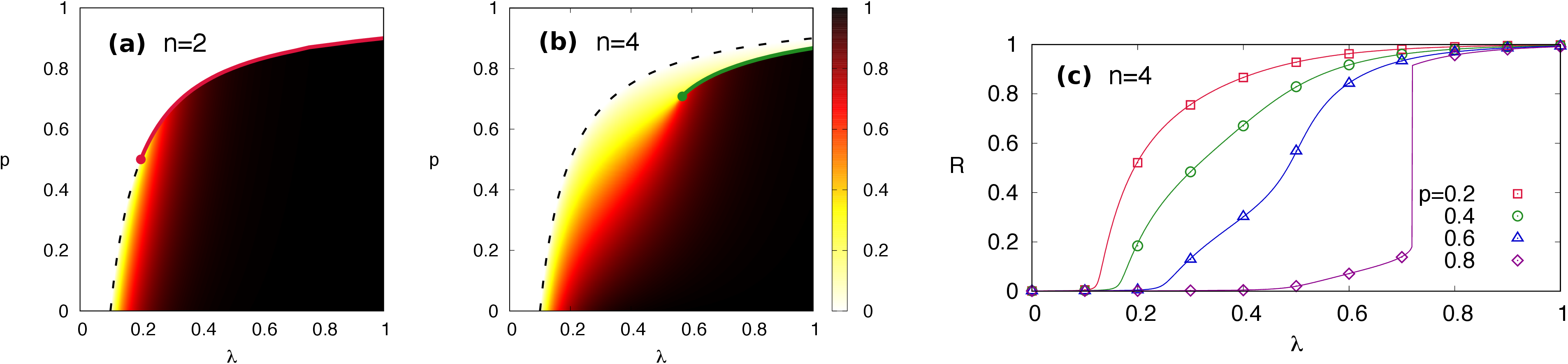}
\caption{
Phase diagram of a generalized contagion model with (a) $n=2$ and (b) $n=4$
showing the final fraction of adopted nodes $R$. Continuous and discontinuous
transition lines are respectively indicated as dashed and solid lines, and
(tri)critical points are indicated by filled circles.
(c) The final fraction of adopted nodes $R$ vs. $\lambda$
with $n=4$ for ER networks with $N=10^5$ and $z=10$,
averaged over $10^4$ independent runs. Numerical simulations (symbol)
and theoretical calculation (line) are shown together.
Error bars are smaller than symbols.
}
\end{figure}

We examine the phase diagram and the fraction of adopted nodes for two specific scenarios
where $n=2$ [Fig.~3(a)] and $n=4$ [Fig.~3(b)] on ER networks with $z=10$.
For $n=2$, when $p<0.5$ a typical continuous phase transition occurs at
$\lambda_1$ [Fig.~3(a)]. But, when more than half of the nodes follow complex contagion
($p>0.5$), the transition between the adoption free phase and the adoption phase becomes
discontinuous. Such discontinuity disappears at a tricritical point
at which $(\lambda_{tc},p_{tc})=(0.2,0.5)$ with $z=10$.
Therefore, for $n=2$ and varying $\lambda$ there is a single transition at $\lambda_1$
either continuous for $p<p_{tc}$ or discontinuous for $p>p_{tc}$.

However, for $n=4$, we find that the continuous and discontinuous transition lines are 
separated [Fig.~3(b)]. To be specific, at a given $p>p_c$ the location of the discontinuous 
transition $\lambda_2$ appears at a value $\lambda>\lambda_1$. 
The size of the jump decreases with decreasing $p$ and
the jump disappears at a critical point $(\lambda_c,p_c) = (0.59,0.71)$.
Therefore above the critical point ($p>p_c$), the size of adopted nodes $R$
abruptly changes from simple adoption phase (low $R$) to complex adoption phase 
(high $R$) at $\lambda=\lambda_2$.
In contrast, below the critical point ($p<p_c$), $R$ changes gradually without 
discontinuity, so that a sharp distinction between simple adoption phase and 
complex adoption phase does no longer exist.

The continuous and discontinuous transitions with $n=4$ and $z=10$
for $p=0.2$, $0.4$, $0.6$, and $0.8$ are shown in Fig.~3(c).
We first note that the theory (line) and numerical simulations (symbol) of $R$ 
for ER networks with $N=10^5$ and $100$ seed nodes show perfect agreement.
In addition, the stark difference between a discontinuous jump for $p>p_c$ and a gradual
increase of $R$ for $p<p_c$ is highlighted.
Note that the fraction of initial seed cannot be negligible in the finite size
networks simulated while it becomes asymptotically small in the thermodynamic limit
$N \rightarrow \infty$.

Moreover, when $p>p_c$, for instance $p=0.8$, the system undergoes a double phase transition with
increasing $\lambda$: a continuous transition from adoption free phase to simple adoption phase, 
followed by a following discontinuous transition between the simple adoption phase
and a complex adoption phase. Recently, multiple transitions in a percolation-type process
have been observed in complicatedly designed networks such as clustered networks~\cite{simon}
and interdependent networks \cite{bianconi}, or with nontrivial percolation protocols
such as explosive percolation \cite{nagler,chen} and asymmetric percolation \cite{allard}.
In this study, however we find a double transition on simple random networks as a result of 
competing contagion processes. It is worthwhile to note that in the limiting case $\lambda=1$ 
our model shares a similarity to the heterogeneous $k$-core percolation which also shows 
a multiple transition \cite{baxter}.

\begin{figure}
\includegraphics[width=\linewidth]{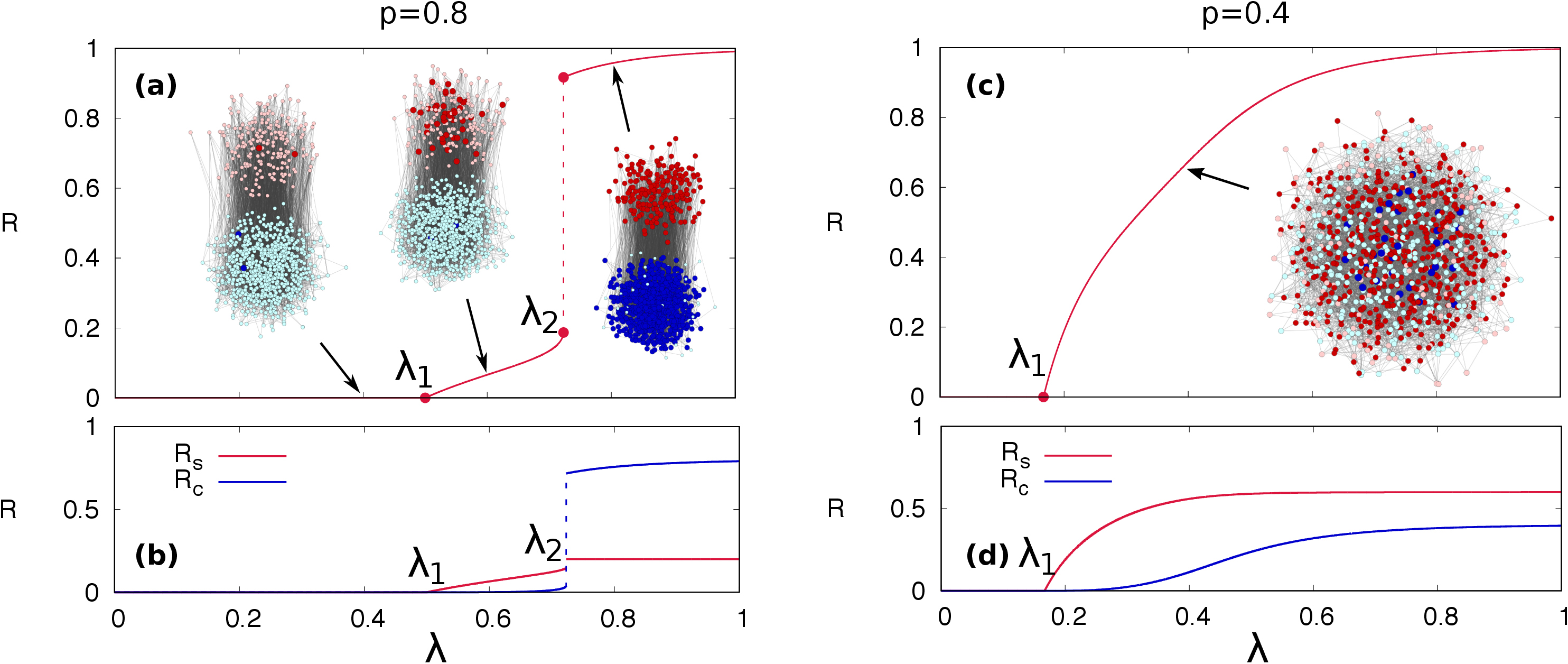}
\caption{
(a) Final fraction of adopted nodes $R$ as a function of $\lambda$ is shown for ER 
networks with $N=10^5$, $z=10$, $n=4$, and $p=0.8$ ($p>p_c$).
Network examples are obtained with the same parameters but for a small network $N=10^3$
for illustration.
Susceptible nodes with simple and complex contagion are indicated by light red
and light blue symbols, respectively. Adopted nodes with simple and complex
contagion are represented as dark red and dark blue, respectively. (b) Final fraction 
of adopted nodes with simple contagion $R_s$ and complex contagion $R_c$ is shown.
(c) $R$ and (d) $R_s$ and $R_c$ are shown for $p=0.4$ ($p<p_c$), disappearing
the distinction between simple adoption phase and complex adoption phase.
}
\end{figure}

\subsection*{Mechanism of double phase transition and mixed phase}

The underlying mechanism of the double phase transition is illustrated in Fig.~4(a),
for ER networks with $z=10$, $n=4$, and $p=0.8$.
In a adoption free phase ($\lambda<\lambda_1$), most nodes, regardless of being 
simple or complex contagion nodes, remain susceptible except initial seeds.
At $\lambda_1$, simple nodes start to become adopted continuously and the
system turns into the simple adoption phase (low $R$). As increasing $\lambda$
above $\lambda_1$, more and more simple nodes become adopted. But, complex
nodes still remain susceptible until $\lambda$ reaches the second transition $\lambda=\lambda_2$.
Therefore, in the simple adoption phase ($\lambda_1<\lambda<\lambda_2$) 
simple contagion nodes are adopted while most of the complex contagion nodes are still
susceptible. As $\lambda$ increases further, at the second transition $\lambda=\lambda_2$
a bunch of nodes with either simple or complex contagion become adopted abruptly.
Thus, in the complex adoption phase ($\lambda>\lambda_2$) most nodes are adopted, leading 
to high $R$. Our numerical simulations for the behavior of the susceptibility of $R$ 
in the limit $\rho \rightarrow 0$ are compatible with a double transition.

The final fraction of adopted nodes with simple contagion $R_s$ and complex contagion
$R_c$ clearly shows the difference between the simple adoption phase and the complex adoption
phase as well as the different mechanisms leading to these two transitions [Fig.~4(b)].
In the simple adoption phase, some of simple nodes become adopted but
complex nodes remain susceptible so that $R_c$ remains zero and $R_s$ has
a finite value. However, in the complex adoption phase, both types of nodes
are adopted, so that both $R_s$ and $R_c$ show a high value after a discontinuous
jump at $\lambda_2$. Note that the maximum of $R_s$ is $0.2$
and that of $R_c$ is $0.8$ because $p=0.8$ in this example.

When $p<p_c$, the discontinuous transition disappears and a single
continuous transition exists at $\lambda_1$ [Fig.~4(c)]. 
As an example, for $p=0.4$ which is less than $p_c=0.71$ both simple nodes and complex nodes 
start to be adopted at $\lambda_1$. And the fraction of adopted nodes with simple $R_s$ and 
complex $R_c$ adoption gradually increases [Fig.~4(d)].
In the network illustration for $\lambda=0.4$ [Fig.~4(c)],
we can observe simultaneously simple nodes and complex nodes that are in the adopted state.
In this mixed phase, simple and complex nodes are strongly interrelated and 
a sharp distinction between a simple and a complex adoption phase is no longer possible.

\section*{Discussion}

In this study, we have proposed a generalized model of contagion processes unifying simple
and complex contagion by introducing an heterogeneous adoptability $Q(\theta)$ together 
with a transmission probability, or link activation probability, that by a simple contagion 
mechanism triggers a cascading complex contagion.
Our model gives rise to diverse phase transitions such as a continuous transition
from adoption free phase to adoption phase, a discontinuous (hybrid) transition
between low adoption and high adoption phase, tricriticality at which two lines 
of the continuous and discontinuous transitions meet, criticality where the discontinuous
transition disappears, and a double transition showing successive occurrence of continuous 
and discontinuous phase transitions when varying the transmission probability $\lambda$.
Specifically, when $n=2$ a continuous transition becomes
discontinuous at a tricritical point. In addition, when $n>2$ continuous and discontinuous
transition lines are separated and two transitions can happen sequentially with
increasing $\lambda$, leading to a double transition.
Our model provides a direction to study general contagion processes
and shows that heterogeneity in agents' response to adoption alters significantly
the consequences of contagion processes. Further studies may be needed to confirm
the finite size effect of the fraction of seed nodes, 
the effect of heterogeneity in network topology, 
and more general adoptability distributions, to name a few.

\section*{Acknowledgements}
We acknowledge financial support from the Agencia Estatal de Investigaci on 
(AEI, Spain) and Fondo Europeo de Desarrollo Regional under project ESOTECOS 
FIS2015-63628-C2-2-R (MINECO/AEI/FEDER,UE).
This work was supported by the National Research Foundation of Korea (NRF)
grant funded by the Korea government (MSIT) (No. 2018R1C1B5044202).

\section*{Author Contributions}
All authors conceived the study,
B. M. performed the research,
all authors analyzed the data,
all authors wrote, reviewed and approved the paper.

\section*{Additional Information}
Competing Interests: The authors declare no competing interests.

\end{document}